\documentstyle{amsppt}
\null
\magnification=\magstep1
\hsize=15.2truecm
\vsize=22truecm
\voffset=0\baselineskip
\par
\noindent
\tolerance=10000
\hfill{hep-th/9801177}
\vskip.15in
\centerline{\bf BEHAVIOUR OF SINGULARITIES
OF THE KERR--NEWMAN}
\centerline{\bf
AND THE KERR--SEN SOLUTIONS BY ARBITRARY BOOST}
\vskip.3in
\centerline{Alexander Burinskii,}
\vskip.15in
 \+&\centerline{\it Gravity Research Group, IBRAE, Russian
Academy of Sciences}\cr
\+&\centerline{\it B. Tulskaya 52, Moscow 113191, Russia; e-mail:
grg\@ibrae.ac.ru}\cr
\vskip.2in
\centerline{Giulio Magli}
\vskip .15in
\+&\centerline{\it Dipartimento di Matematica del Politecnico di Milano,}\cr
\+&\centerline{\it Piazza Leonardo Da Vinci 32, 20133 Milano, Italy;
e-mail magli\@mate.polimi.it}\cr
\vskip.3in
\midinsert
\baselineskip=14pt plus.1pt
\vskip 4 truecm
\par
\vskip.1in

The behaviour of the singularities of the rotating black holes
under an arbitrary boost is considered
on the basis of a complex representation of the Kerr theorem.
We give a simple algorithm allowing to get
explicit expressions for the metric and the position of the
singularities for
arbitrary direction and magnitude of the boost,
including the ultrarelativistic case.
The non-smoothness of the ultrarelativistic limit is discussed.
The Kerr-Sen BH-solution to low energy string theory is also analyzed.

\endinsert
\vfill\eject
\baselineskip=17pt

\bigskip

\def\b{\bar}

\def\d{\partial}
\def\D{\Delta}

\def\l{\lambda}

\def\m{\mu}
\def\n{\nu}

\def\q{\b q}

\def\t{\tau}
\def\x{\phi}

\def\~{\tilde}
\def\h{\eta}
\def\z{\zeta}
\def\Z{{\b\zeta}}
\def\Y{{\b Y}}

\def\`{\dot}

\par\noindent
{\bf 1. INTRODUCTION}
\par
\bigskip

Recently,
the problem of finding
the ultrarelativistic limit of exact particle-like
solutions of the Einstein field equations
received considerable attention,
especially in connection with some non-trivial
gravitational effects which are expected to occur in the interparticle
interactions at extreme energies due to the appearance of gravitational
shock waves [1-8].
\par
First results in this field were obtained by Aichelburg and
Sexl [1], who considered the ultrarelativistic boost of the Schwarzschild
solution to analyze the behaviour of the gravitational field of a massless
point particle in the light-like limit.
\par
A similar treatment for the Kerr geometry, which can be considered
as a model of a spinning particle in general relativity, has to take
into account the orientation of the angular momentum with respect to
the boost [4-8].
\par
There are three different physical situations
connected with boosted BH-solutions.
The first one is the original Aichelburg--Sexl
problem of application of such solutions to
describe the gravitational field of light--like particles
with or without spin.
The second application
consists in modelling the gravitational field
of elementary particles with finite rest mass
under the boost, and it is connected with an analysis of possible effects
generated in relativistic collisions.
Finally, there are astrophysical applications,
namely boosting black holes.
In this case also the behaviour of the horizon and of the ergosphere
under the boost are of interest.

The analysis of the boosted Kerr solution [4-8] exhibits
some difficulties in the interpretation of the results in the limiting,
ultrarelativistic case.
In particular, there are technical difficulties
due to the absence of smooth
ultrarelativistic limits, as well as an ambiguity in performing the limits
when more than one parameter is
involved in the limiting procedure simultaneously
(for example, the parameters $m$ and $a$ in the non--charged Kerr case).
\par
In any of the above cited approaches,
the boosted Kerr solution is given by approximate
expressions so that one cannot obtain an invariant
description of the behaviour of the singularity under the boost.
We propose here a different method of description
of the boosted Kerr solution based on the Debney,
Kerr and Schild formalism [9] (DKS) and on the Kerr theorem [9-13].
\par
The advantage of this approach relies in the possibility
of obtaining exact, {\it explicit} expressions for the
metric and its singularities
in the case of an {\it arbitrary} boost,
namely a boost with an arbitrary orientation with respect to
the angular momentum.
In fact,
being represented in the Kerr-Schild form, the boosted Kerr
metric can be linked with an auxiliary Minkowski space having a ``rigid''
coordinate system. This allows us
to represent shock waves and singularities in asymptotically flat
Cartesian coordinates.

\bigskip
\par\noindent
{\bf 2. THE DKS-FORMALISM AND THE KERR THEOREM}
\bigskip

In our notations we follow the work of Debney, Kerr and Schild [9].
All the BH-solutions in Einstein's gravity
can be described by the simple Kerr-Schild metric
$$
g_{\m\n} =
\h_{\m\n} + 2 h e^3_{\m} e^3_{\n}, \eqno(2.1)
$$
where $\h_{\m\n}$ is the metric of an auxiliary Minkowski space $M^4$
with signature $(+++-)$ and Cartesian coordinates $t,x,y,z$.
For a non-rotating BH the scalar function $h$ has the form
$$
h= m/r - e^2 / 2r^2,
$$
where $m$ and $e$ are the mass and the charge of the BH.
The vector  $e^{3\m} = (1, \vec k)$ is a field of principal null directions
which is spherically symmetric ($\vec k=(x,y,z)/r$ in the auxiliary Minkowski
space with metric $\h_{\m\n}$).
In the case of rotating BH-solutions the metric is still of
the Kerr-Schild form but its twisting structure
is determined by a different null congruence $e^3$ and by a modification
of the radial coordinate.

In null coordinates
$$
\eqalign{2^{1\over2}\z &= x+iy ,\qquad 2^{1\over2} \Z = x-iy ,\cr
2^{1\over2}u & = z + t ,\qquad 2^{1\over2}v = z - t,}  \eqno (2.2)
$$
the null vector $e^3$ can be expressed
via a scalar function $Y(x)$ in the following way:
$$
e^3 = du+ \Y d \z  + Y d \Z - Y \Y d v.\eqno(2.3)
$$

The determination of $e^3$
is possible since
the principal null congruences of rotating BH solutions
are geodesic and shear-free, and
the {\it Kerr Theorem} [9-13]
gives a rule to construct all such congruences:
an arbitrary,
geodesic shear-free null congruence in
Minkowski space is defined by a function $Y (x)$ which is a solution of the
equation
$$
F  = 0 , \eqno(2.4)
$$
where
$F (\l_1,\l_2,Y)$ is an arbitrary analytic
function of the {\it projective twistor coordinates}
$$ \l_1 = \z - Y v, \qquad \l_2 =u + Y \Z, \qquad Y\ .\eqno(2.5)$$

A consequence of the Kerr Theorem is also the expression for
the complex radial coordinate
$$\tilde r \equiv PZ^{-1} =dF/dY, \eqno(2.6) $$
which characterizes ``{\rm dilatation} $+i$ {\rm twist }'' of the
congruence.
Correspondingly, the singular regions of the metrics
are defined by the system of
equations
$$
F=0, \qquad
d F / d Y =0.  \eqno(2.7)
$$
The BH-solutions belong to a class of metrics for
which the singularities are contained in a bounded region of space.
In this case the equation $F=0$ can be solved in explicit form.
Moreover, in this case there exists a complex representation of the
function F in which the congruence is defined by an effective "source"
moving in complex Minkowski space $CM^4$ along a complex world line.
Such a complex representation was initially suggested by Lind and Newman
[14,15] in the Newman-Penrose formalism.
The field $e^3$ can be used as one of the vectors of
null tetrad $ e_1, e_2, e_3, e_4 $
satisfying
$$
g_{ab}=
e_a^\m e_{b\m} =\left(\matrix 0&1&0&0\cr1&0&0&0\cr 0&0&0&1\cr0&0&1&0
\endmatrix \right) = g^{ab}, \eqno(2.8)
$$
($e^3, e^4$ are real null
vectors, $ e^1, e^2 $ are complex conjugates).
The null tetrad $e_a^\m$ can be completed
as follows:
$$
\eqalign{
e^1 &= d \z - Y d v; \cr e^2 &= d \Z - \Y d v; \cr e^4 &=  d v - h e^3. }
   \eqno(2.9) $$ The inverse tetrad has the form $$\eqalign{ \d_1 &= \d_\z
 - \Y \d_u ; \cr \d_2 &=  \d_\Z - Y \d_u ; \cr \d_3 &=  \d_u - h \d_4 ;
\cr \d_4 &=  \d_v + Y \d_\z + \Y \d_\Z - Y  \Y \d_u .  } \eqno(2.10)
$$

The function $h$ of the Kerr-Newman solution has the
form
$$ h= m (Z +\bar Z)/P^3 -e^2/(Z \bar Z), \eqno(2.11) $$
while the
electromagnetic field can be obtained from the potential
$$ A= -e(Z+\bar Z)e^3/(2P^2). \eqno (2.12)$$

\bigskip
\par\noindent
{\bf 3. WEAK STATIONARIETY AND CONGRUENCES HAVING
SINGULARITIES CONTAINED IN  A BOUNDED REGION}
\bigskip

The null congruence with tangent $ e^3 $ is stationary in $M^4$ if
$\partial _t e^3 =0$. However, in general one can also consider
a "weak nonstationariety" corresponding to the fact that
the stationariety can be
restored by a Lorentz transformation. In this case
there exists a real time-like vector field K such that
$$
K Y = K \bar Y = 0, \eqno (3.1)
$$
and consequently $K e^3 =0$.
The congruences stationary
in this weak sense and having singularities contained
in a bounded region have been considered in [16-18].
In this case the function $F$ must be at most quadratic in $Y$,
$$
F \equiv a_0 +a_1 Y + a_2 Y^2 + (q Y + c) \l_1 - (p Y + \q) \l_2,\eqno(3.2)
$$
where the coefficients $ c$ and $ p$ are real constants
and $a_0, a_1, a_2,  q, \q, $  are complex constants.
The solutions of the equation $ F = 0$ and the equations for the
singularities can be found in this case in explicit form.
The solution $Y(x)$ of the equation $ F=0 $ satisfies the weak
stationariety condition (3.1) if
$$ K= c \d_u + \bar q \d_\z + q \d_\Z - p \d_v . \eqno(3.3)$$

In the papers [16,17] another,
equivalent form of  $ F $ was suggested.
This form
allows to represent the parameters of the
function $F$ and the vector field
$K$ as retarded-time fields starting
from an ``effective'' complex world line
$x_0^\mu(\tau)$ depending from a complex time parameter $\tau$.
This form is the following
$$
F \equiv (\l_1 - \l_1^0) K \l_2 - (\l_2 -\l_2^0) K \l_1.\eqno(3.4)
$$
Here the
twistor components with zero indices
$$ \l_1^0 (\t)= \z_0(\t) - Y v_0(\t), \qquad
 \l_2^0(\t) =u_0(\t) + Y \Z_0(\t),\eqno(3.6) $$
denote the values on the points of the complex world-line
represented in null coordinates
$\x_0 (\t)= (\z_0, \Z_0, u_0 ,v_0)$ ($\Z_0$ and $\z_0$ are not
necessarily complex conjugates).
The vector $K$ can be expressed
in the form
$$ K(\t) = \`x_0^\m(\t) \d_\m, \eqno(3.7)$$
where the dot denotes $\d_\tau$.

The Kerr congruences with weak nonstationariety are determined by straight
analytic world lines with constant 3-velocity $\b v$:
$$ x_0^\m (\t) = x_0^\m (0) + \xi^\m \t; \qquad \xi^\m = (1,\b v),
\eqno(3.8)$$
correspondingly, the vector $K=\xi^\m \d_\m$  is a constant Killing vector
of the solutions.
The form (3.4) has the remarkable property that, in spite of an explicit
dependence of the parameters of the function $ F$ in (3.3) on $\t$, this
dependence is absent really, since in consequence of the relations
$$\l_1^0 (x_0(\t)) = \l_1^0 (x_0(0)) + \t K \l_1,\quad \l_2^0 (x_0(\t)) =
\l_2^0 (x_0(0)) + \t K \l_2,  \eqno (3.9)$$
the terms proportional to $ \t$ cancel.
Therefore the expressions (3.1) and (3.4) are equivalent.

The relation (3.4) is very convenient
in order to obtain explicit representation of the
congruences of the boosted Kerr solution.
By writing the function F in the form $$ F = A
  Y^2 + B Y + C, \eqno (3.10)$$ where  $$\eqalign { A &= (\Z
  - \Z_0) \`v_0 - (v-v_0) \`\Z_0 ;\cr B &
= (u-u_0) \`v_0 + (\z - \z_0 ) \`\Z_0
  - (\Z - \Z_0) \`\z_0 - (v - v_0) \`u_0 ;\cr C
&= (\z - \z_0 ) \`u_0 - (u -
u_0) \`\z_0, } \eqno(3.11) $$ one can find two explicit solutions for the
function $Y(x)$
 $$ Y_{1,2} = (- B \pm \D )/2A, \eqno(3.12)$$
 where $ \D = (B^2 - 4AC)^{1/2}.$

On the other hand differentiating $F=0$ and using (2.6) one finds
 $$ Y = - (B + PZ^{-1})/2A, \eqno (3.13) $$ and consequently
$$PZ^{-1} = \mp \D. \eqno (3.14)  $$
One can find also  $$ P = \`x_o^\m(\t) e^3_\m . \eqno (3.15).$$

The field $e^3$ can be normalized by introducing $l^\mu = e^{3 \mu}/P$ so
that ${\dot x}_0^\m l_\m =1, $ that yields the following
form of the Kerr-Newman metric
 $$g_{\m\n} =\h_{\m\n} + [m({\tilde r}^{-1} + {\bar {\tilde r}}^{-1})
  - e^2 (\tilde r {\bar {\tilde r}})^{-1}] l_{\m} l_{\n}. \eqno (3.16) $$
where the complex radial coordinate $\tilde r\equiv PZ^{-1}$ is given
by the expression (3.14)
or can be represented in the form
$$\tilde r = -dF/dY = -B - 2A Y . \eqno (3.17) $$
It is convenient to represent $\tilde r$
as a sum of the real radial distance $r$ and an
angular coordinate  $\tilde r=  r +i a \cos \theta $.
Then the equation (3.14)  fixes the relation between the polar coordinates
$r, \theta, \phi$ and the null Cartesian coordinates through the
expressions (3.11) for the coefficients $A,B,C$.

\bigskip
\par\noindent
{\bf 4. BEHAVIOUR OF SINGULARITIES OF THE KERR-NEWMAN SOLUTION BY THE BOOST}
\bigskip

In the ``gauge'' $x_0^0 =\tau$
the complex world line (3.8) can be represented as
$x_0^\mu (\tau) = \{ \tau, \vec x_0 (0)+ \vec v \t \}$.
The complex initial displacement can be decomposed as
$\vec x_0 (0) = \vec c + i\vec d$, where $\vec c $ and $\vec d$
are real 3-vectors with respect to
the space O(3)-rotation. The real part $\vec c$
defines the initial shift of the solution, and the imaginary part $\vec d$
defines the size and the position of the
singular ring as well as the corresponding angular
momentum.  It can be easily shown that in the rest frame, when $\vec V=0,
\quad \vec d =\vec d_0 $, the singular ring lies
in the plane orthogonal to
$\vec d$ and has a radius $a=\vert \vec d_0 \vert $. The corresponding
angular momentum is $\vec J = m \vec d_0.$
\par
In the case of a boost
orthogonal to the direction of
$\vec d$, this vector is not altered by Lorentz contraction
($\vec d=\vec d_0$, $\vert \vec d \vert =a$),
while if $\vec d$ and $\vec V$ are collinear we have
$$
\vec d_0=\vec d/\sqrt{1-\vert\vec V \vert^2}\ . \eqno(4.1)
$$
This shows that the parameter $a$ coincides with its rest value $a_0$ if
$\vec d $ and $\vec V$ are orthogonal, while
$$
a_0=a/\sqrt{1-\vert\vec V \vert^2}\ , \eqno(4.2)
$$
if $\vec V$ and $\vec d$ are collinear.
\par
In order to calculate the parameters $A,B,C$ it is convenient
to express the complex world line in null coordinates
$$
\eqalign{2^{1\over2}\z _0 &= x_0+iy_0 ,
\qquad 2^{1\over2} \Z _0 = x_0-iy_0 ,\cr
2^{1\over2}u_0 & = z_0 + t_0 ,\qquad 2^{1\over2}v_0 = z_0 - t_0}\ .
\eqno (4.3)
$$
The Killing vector of the solution will then be
$$
\xi^\mu = 2^{-1/2}\{\` u_0-\`v_0, \` \z_0 +\`\Z_0, -i( \` \z_0 +\`\Z_0),
\` u_0+\`v_0 \},
$$
while the functions  $P$ takes the form
$$
P= e^3_\mu {\dot x}_0^\mu =
\` u_0 + \Y \` \z_0  + Y  \` \Z_0 - Y \Y  \`v_0 .\eqno (4.4)
$$
\par
The complex radial coordinate $\tilde r\equiv PZ^{-1}$ is given
by the expression (3.4).
As for the unboosted Kerr solution,
one can represent $\tilde r$
as a ``sum'' of the real radial distance $r$ and an
angular coordinate.
Then equation (3.4) can be used to fix
the relation between the polar coordinates
$r, \theta, \phi$ and the null Cartesian coordinates (4.3) through the
expressions (3.3) for the coefficients $A,B,C$.
Due to the formula (3.17),
the singular regions are defined by the zeros of the
function $\tilde r.$
In what follows,
we present some examples of boosted Kerr solutions
and then discuss the general features exhibited by them.
\par
\bigskip
{\bf Example I.}
\par
\bigskip

Spinning particle moves with speed of the light in
the positive direction of the $z $- axis, 3-vector
${\vec d}= (0,0,a)$ is also directed along the $z$-axis.
We have the following  coordinates of complex world line

$x_0^0 (\tau) \equiv \tau$,
$ z_0 (\t) =ia + \tau, \quad x_0(\tau)=y_0(\tau)=0.$
\par
In the null coordinates it gives
$$\sqrt{2} u_0= z_0+\tau=ia+2\tau; \quad \sqrt{2} v_0= z_0-\tau =ia, \quad
\z_0=\Z_0=0,\eqno(I.1)$$
that yields
\par
$\`u_0= \sqrt{2}, \quad \`v_0 =0, \quad
\` \z_0 =\`\Z_0 =0,  $
and
$$\eqalign{u-u_0 &= (z-ia + t - 2\tau)/\sqrt{2},\cr
v-v_0 &= (z-ia -t )/\sqrt{2},\cr
 \z-\z_0 &=\z, \qquad\Z-\Z_0=\Z}. \eqno(I.2)$$

Coefficients $A,B,C$ calculated from (3.6) will be

$$A=0;\quad B=t-z+ia;\quad C= x+iy. \eqno(I.3)$$

As a result the function $F$ acquires the form
$F=x+iy - Y (z-ia -t)$, and solution of the equation $F=0$ is
$$Y= (x+iy)/(z-ia-t).\eqno(I.4)$$
The function $$\tilde r = - dF/dY = z-ia-t.
 \eqno(I.5)$$
One can see that there is no singularity if $a\ne 0$ since there is no real
solutions to the system of equations $F=F_Y=0 $.

On the other hand, setting $a=0$ we have got the case of spinless particle,
and a moving singular plane  which is placed at $z=t.$ Therefore there is
no smooth limit by $a\rightarrow 0$.

\bigskip
\noindent
{\bf Example II.}
\par
\bigskip

The motion with speed of the light in
the positive direction of the $x$- axis, orthogonal to the 3-vector $\vec d$
which defines the direction and the value of the
angular momentum ${\vec J}= m(0,0,a)$,
$a=\vert \vec d \vert$.
We have the complex world line $ x_0 (\t) = \tau,\quad y_0(\tau)=0,\quad
z_0 (\t) =ia,\quad t_0=\tau.$
Correspondingly, the world line in the null coordinates is
\par
$$\sqrt{2} u_0=ia+\t,\quad\sqrt{2} v_0=ia-\t,\quad\sqrt{2} \z_0=\t,\quad
\sqrt{2} \Z_0=\t;
 \eqno(II.1)$$
and the velocities are
\par
$\sqrt{2} \dot u_0=1,\quad\sqrt{2} \dot v_0=-1,\quad\sqrt{2} \dot \z_0=1,
\quad \sqrt{2} \dot {\bar \z_0}=1.$
\par
We have also
$$\eqalign{\sqrt{2}(u- u_0) &=z+t -ia-\t,\quad\sqrt{2}(v- v_0)=z-t-ia+\t,\cr
\sqrt{2} (\z-\z_0)&=x+iy -\t,\quad
\sqrt{2} (\bar \z-\bar \z_0)=x-iy-\t.}\eqno(II.2)$$

Coefficients $A,B,C$ take the form
$$A= (-x+iy -z+t+ia)/2;\qquad B=ia+iy -z;\qquad C= (x+iy -z-t+ia)/2.
 \eqno(II.3)$$
The function $\tilde r\equiv PZ^{-1}$ takes the form
$$ PZ^{-1} = -d F/dY = x-t. \eqno(II.4)$$
There is therefore a moving singular plane placed at  $x=t.$
The function $Y$ will be
$$Y= (dF/dY -B)/2A =(x+iy - z -t+ia)/(x-iy +z -t-ia).
\eqno(II.5)$$
\par
\bigskip
\noindent
{\bf Example III.}
\par
To understand better the absence of smooth limit in the example I we
consider here an intermediate case with a boost with a
speed
$v=\alpha c$,  $\alpha\le 1$ in the positive direction of the $z $- axis,
and then we will consider the limit $ \alpha\rightarrow 1 $.
As in the example I, the 3-vector $\vec d = (0,0,a)$  is directed along the
$z$-axis. We have the complex world line
$$ x_0 (\t) = y_0(\t)=0,\quad z_0 (\t) =ia+\alpha\t,
\quad t_0=\tau.
\eqno(III.1)$$
In null coordinates we have
$\sqrt{2} \dot u_0=1+\alpha, \quad \sqrt{2} \dot v_0=-1+\alpha,\quad
\sqrt{2} \dot \xi_0=0,
\quad \sqrt{2} \dot {\bar \xi_0}=0.$
\par
It yields
$$\eqalign{\sqrt{2}(u- u_0) &=z+t -ia-(\alpha+1)\t,\quad
\sqrt{2}(v- v_0)=z-t-ia+(1-\alpha)\t,\cr
\sqrt{2} (\xi-\xi_0) &=x+iy,\quad
\sqrt{2} (\bar \xi-\bar \xi_0)=x-iy.}
\eqno(III.2)$$
Coefficients $A,B,C$ will be
$$A= -(x-iy)(1-\alpha)/2;\quad B=ia -z + \alpha t;\quad
C= (x+iy)(1+\alpha)/2.
\eqno(III.3)$$
The expression for complex radial distance (3.9) yields

\par
$$ {\tilde r}^2 = B^2 -4AC= (z-\alpha t)^2 +
(1-\alpha^2) (x^2+y^2) -a^2 -2ia (z-\alpha t).\eqno(III.4)$$
Like to the standard Kerr solution one can represent the complex radial
coordinate $\tilde r$ as a sum of the real radial distance $r$ and an
angular coordinate  $\tilde r=  r +i a \cos \theta $. Then, selecting the
real and imagine parts of the expression (III.4) one obtains the following
relations between the polar coordinates $r, \theta, \phi$ and  Cartesian
coordinates of the auxiliary Minkowski space

$$\eqalign {x+iy &= (r+ia) e^{i\phi} \sin \theta /\sqrt{1-\alpha^2},\cr
 z-\alpha t &= r \cos \theta.} \eqno(III.4)$$

For the case $\alpha=0$ this coincides with the coordinate relations of the
standard Kerr solution. Setting $ r=\cos \theta =0$ we obtain the equation
of singular ring
$$x^2+ y^2 = a^2 /(1-\alpha^2),\qquad  z-\alpha t = 0. \eqno(III.5)$$

It may be seen that size of the ring grows by the increasing of $\alpha $,
and in the  limiting case $\alpha=1$ the singularity
is placed on  infinity.

The cause of this effect is the above mentioned relation (4.2)
$a_0=a/\sqrt{1-\alpha^2}$. The increasing of singular ring by
$\alpha=v/c \rightarrow 1$ is a seeming effect connected with using the
parameter $a$ instead of its rest value $a_0$.
Being to expressed via the rest value singular region takes
the form of moving ring of constant radius $a_0$, however, if we consider
a light-like particle its rest mass is infinitely small and singularity
is to be placed on infinity.
\par
\bigskip
\noindent
{\bf Example IV.}
\par
Intermediate case clarifying the limiting result of example II.
The boost with  a speed
$v=\alpha c$, $\alpha\le 1$ in the direction $x$,
orthogonal to  direction of angular momentum  ${\vec d}= (0,0,a)$.
\par
We have the complex world line
$$ x_0 (\t) = \alpha \t, \quad y_0(\t)=0, \quad z_0 (\t) =ia,\quad t_0=\tau;
\eqno(IV.1)$$
In the null coordinates it yields
\par
$\sqrt{2} \dot u_0=1, \quad\sqrt{2} \dot v_0=-1,
\quad\sqrt{2} \dot \xi_0=\alpha, \quad\sqrt{2} \dot {\bar \xi_0}=\alpha.$
$$\eqalign{\sqrt{2}(u- u_0) &=z+t -ia-\t,\quad
\sqrt{2}(v- v_0)=z-t-ia+ \t,\cr
\sqrt{2} (\xi-\xi_0) &=x+iy-\alpha \tau,\quad
\sqrt{2} (\bar \xi-\bar \xi_0)=x-iy -\alpha \tau.}
\eqno(IV.2)$$
Coefficients $A,B,C$ will be the following
$$A= -(x-iy)/2 -(z-t-ia)\alpha /2;\quad B=ia-z+ i y \alpha ;
\quad C= (x+iy)/2 -\alpha(z+t-ia)/2.
\eqno(IV.3)$$
>From the equation  (4.4) we obtain
$$ \tilde r ^2 = B^2 -4AC= (x-\alpha t)^2 + (1-\alpha^2)[y^2+ (z-ia)^2].
\eqno (IV.4)$$
Representing $\tilde r=r+i a \sqrt{1-\alpha^2} \cos\theta$ and selecting
the real and
imaginary parts of (IV.4) one obtains the following
coordinate relations which generalize corresponding relations of
the stationary Kerr solution
\par
$$\eqalign{(x-\alpha t)/\sqrt{1-\alpha^2} + iy &= (r/\sqrt{1-\alpha^2} +ia)
e^{i\phi}\sin\theta,\cr
z &= r \cos\theta/\sqrt{1-\alpha^2}.} \eqno (IV.5)$$
Singular region $r=\cos\theta=0$ will be
$$z=0;\qquad  (x-\alpha t)^2 + (1-\alpha^2) y^2 = a^2 (1-\alpha^2).
\eqno(IV.6)$$
This  is a moving ring placing in the $z=0$ plane. It is oblate in $x$
direction with the Lorentz factor $\sqrt{1-\alpha^2}$.

Therefore, in the limit $\alpha=1$ singular region will be moving
segment of the line $z=0, x=t;\quad -a \le y \le a,$ which is parallel to
the $y$-axis.
One can mention that this limit is also non-smooth.

 So, setting $\alpha=1$ in the equations (IV.4),(IV.5) we come to  equation
 $x=t$, coordinate $y$ is not restricted, however, the coordinate $z$ is
 indefinite  $\sim 0/0$. It will be equal to
zero if we  set first $r=\cos \theta=0$ corresponding the singular region
and then take the limit $ \alpha\rightarrow 1$, however it is equal to
infinity if the limit $\alpha=1$ is taken first.
\par
\bigskip
\noindent
{\bf Example V.}
\par
\bigskip
We finally consider the general case in which
the value of the velocity
is arbitrary as well as its direction
with respect to the angular momentum.
Without lost of generality,
we can consider
the boost performed with a parameter
$\alpha $ in the $z$-direction $(\alpha=v_z/c)$, and a parameter $\beta$ in
the $x$-direction $(\beta=v_x/c)$, while
the angular momentum is defined by  ${\vec d}= (0,0,a)$.
Denoting
$w^2=\alpha^2 +\beta^2$
the following general formula for the coordinate relations
can be easily obtained:
$$
\eqalign{(x-\beta t)\sqrt{1-\alpha^2} + iy \sqrt{1-w^2} &=
(r +i a\sqrt{1-\beta^2}) e^{i\phi}\sin\theta,\cr
z- \alpha t &= - r \cos\theta/\sqrt{1-\beta^2}.}  \eqno(V.1)
$$
The singular region $r=0, \cos\theta=0$ is placed on the plane
$z=\alpha t$ and is described by
$$
\qquad  (1-\alpha^2)(x-\beta t)^2 + (1-w^2) y^2 = a^2 (1-\beta^2).
\eqno(V.2)
$$
The singularity is a moving ring
distorted in the $x$ direction by a
factor $\sqrt{{(1-\beta^2)}/{(1-\alpha^2)}}$ and in the $y$ direction
by a factor $\sqrt{{(1-\beta^2)}/{(1-w^2)}}$.
The ultrarelativistic limit corresponds to $w=1$
and the singular region is a couple
of straight lines parallel to the $y$ axis.
\par
\bigskip
\noindent
Therefore,
we can conclude that the non--smoothness and the non--commutativeness
of the limiting procedure
is a {\it general}
feature of the boosted Kerr solutions.
\medskip
The main consequences of the considered examples are the non-smoothness
and the non-commutativeness of the limiting procedure as well as
an unexpected
behaviour of the Kerr singular ring which is connected with
the definition of the parameters of the solution after
the boost and shows that such parameters must be ``renormalized''
by the boost.

It is interesting also to observe from the coordinate relations
that the coordinate $r$ is 'scaled' by the boost with respect to the
asymptotically flat Cartesian coordinates $x,y,z$;
as a consequence the region of small
values of $r$ (and big values of $h$) is stretched to big values of
the $x,y,z$-coordinates,
this is the origin of the shock waves in the ultrarelativistic limit
since the region of big $h$ can be very far from the "centre" of the
solution.
\par
In astrophysical applications,
the behaviour of the horizon and of
the ergosphere after the boost
also has a physical interest.
It can be easily shown
that in
the above suggested coordinates $r$ and $\theta$
the horizon as well as the ergosphere
are simply given by the known formulae for the
Kerr case where the mass parameter $m$ must be
scaled by the Lorentz factor.

\par
\bigskip
\par\noindent
{\bf 5. BOOST OF THE KERR-SEN SOLUTION}
\bigskip
\par

Recently, rotating BH-solutions
received attention also in string theory.
The Kerr-Sen BH-solution is a generalization of the Kerr solution to
low energy string theory [19] (or to axion-dilaton gravity).
We would like to show that the
above formalism is also applicable to the Kerr-Sen solution.
The metric of the Kerr-Sen BH may be written in the  form [20]
$$
 ds^2_{dil}=2e^{-2(\Phi - {\Phi}_0)} {\tilde e}^1
{\tilde e}^2 +
2{\tilde e}^3 {\tilde e}^4, \eqno(5.1)
$$
where
$$
{\tilde e}^1 =(PZ) ^{-1} dY ,
\qquad{\tilde e}^2 =(P\bar Z) ^{-1} d \bar Y ,
\eqno(5.2)
$$
$$
{\tilde e}^3 = P^{-1} e^3 ,
\eqno(5.3)
$$
$$
{\tilde e}^4 = dr + iaP^{-2}(\bar Y dY - Y d \bar Y) +
 (H_{dil} -1/2)e^3,
\eqno(5.4)
$$
and

$$
H_{dil}  = M r/ \Sigma_{dil};\quad
\Sigma_{dil}=e^{-2(\Phi - {\Phi}_0)}(Z{\bar Z})^{-1};
\eqno(5.5)
$$
$$
e^{-2(\Phi - {\Phi}_0)} = 1 + (Q^2/2M)(Z+{\bar Z});
\qquad Z^{-1} \equiv \tilde r.
\eqno(5.6)
$$
The field of principal null directions is $\tilde e^3$.
Following eq.(6.1) of [9] this tetrad is related to the Kerr-Schild
tetrad (2.3),(2.9) as follows
$$
{\tilde e}^1 =e^1 - P^{-1} P_{\bar Y} e^3, \qquad
{\tilde e}^2 =e^2 - P^{-1} P_{Y} e^3,
\eqno(5.7)
$$
$$
{\tilde e}^3= P^{-1} e^3,
\eqno(5.8)
$$
$$
{\tilde e}^4 = Pe^4_{dil} + P_{ Y} e^1 + P_{\bar Y} e^2
-P_{Y}P_{\bar Y}P^{-1}e^3.
\eqno(5.9)
$$
Therefore the Kerr-Sen metric (5.1) may be reexpressed
in the form containing the Kerr-Schild tetrad $e^a$,
the dilaton factor
$e^{-2(\Phi - {\Phi}_0)} $, and a ``deformed'' function
$$
 H_{dil} = h e^{2(\Phi - {\Phi}_0)}
$$
instead of the function $h$.
\par
It was shown in [20] that the Kerr-Sen metric is of type I
contrary to the Kerr solution which is
type D.
However one of the principal null
directions $ e^3$ of the Kerr and the Kerr-Newman solutions survives in the
Kerr-Sen solution and retains the property of being geodesic
and shear free. It means that the Kerr theorem is applicable to this
solution, since it has just the same principal null congruence
and positions of caustics.
Therefore the above analysis
can be extended to the Kerr-Sen solution.

\bigskip
\par\noindent
{\bf 6. CONCLUSIONS}
\bigskip
\par

We discussed here a method allowing to describe in {\it explicit}
form the metric and
the behaviour of the singular region of the Kerr solution
under arbitrary boost and with arbitrary orientations of angular momentum.
In particular, we have shown that
the Kerr theorem automatically allows to obtain
an asymptotically flat coordinate system and the equations
describing
the singularities in these coordinates.
These results throw
some light on the somewhat mysterious ``standard'' procedure
commonly used to obtain shock waves metrics.
In fact this procedure gives only
approximate expressions before taking the ultrarelativistic limits [5-8].
Of course, the
tail of the shock wave (logarithmic term) cannot be obtained using the
present method, because first of all we work always with vacuum, singular
solutions of the Einstein field equations.
To obtain the profile of the wave located on the delta--like singularity
of the metric, one must first re--interpret the metric itself as being
created by a singular distribution of matter on an extended manifold.
\par
Our results are not very encouraging as far as the physical
content of all such ultrarelativistic solutions is concerned.
In fact, we obtained a quite general picture of non-smoothness and
non-commutativeness of
the limits $a \rightarrow0$, $v \rightarrow 1$ and $r\rightarrow 0$.
The absence of a smooth limit
explains the well known fact
that the limiting solution belongs to a completely
different class
with respect to the starting one:
it is type $N$ and not type $D$ as
the original Kerr solution,
it is not asymptotically flat
and has another group of symmetry.
\vfill\eject

\bigskip
\par\noindent
{\bf ACKNOWLEDGEMENTS}
\bigskip

The authors acknowledge Elisa Brinis Udeschini for
many interesting discussions. One of us, A.B., is grateful to P. Aichelburg
for useful discussions and hospitality at the Vienna University.

\bigskip
\vfill
\eject
\newpage
\par\noindent
{\bf REFERENCES}
\frenchspacing
\medskip
\baselineskip=15.5pt

\item{[1]}
P.C.Aichelburg and R.U. Sexl, Gen. Rel.Grav. {\bf 2} (1971) 303
\item{[2]}
T. Dray and G. 't Hooft,
Nucl. Phys. B {\bf 253} (1985) 173
\item{[3]}
M. Fabbichesi, R. Pettorino, G. Veneziano and G.A. Vilkovisky,
Nucl. Phys. B {\bf 419} (1994)147
\item{[4]}
H. Balasin and H. Nachbagauer, Class. Quantum Grav. {\bf 12} (1995) 707
\item{[5]}
H. Balasin and H. Nachbagauer, Class. Quantum Grav. {\bf 13} (1996) 731
\item{[6]}
V. Ferrari and P. Pendenza, Gen. Rel. Grav. {\bf 22} (1990) 1105
\item{[7]}
C.O.Luosto and N. Sanchez, Nucl. Phys. B {\bf 355} (1991) 231
\item{[8]}
C.O.Luosto and N. Sanchez, Nucl. Phys. B {\bf 383} (1992) 377
\item{[9]}
G.C. Debney, R.P. Kerr and  A. Schild, J. Math. Phys. {\bf10} (1969) 1842.
\item{[10]}
R. Penrose, J. Math. Phys. {\bf8} (1967) 345.
\item{[11]}
D.Cox and E.J. Flaherty, Comm. Math. Phys. {\bf 47} (1976) 75.
\item{[12]}
D.Kramer, H.Stephani, E. Herlt, M.MacCallum, {\it Exact Solutions
of Einstein's Field Equations,} Cambridge Univ. Press, Cambridge 1980.
\item{[13]}
E.Brinis Udeschini and G. Magli,
J. Math. Phys. {\bf 37} (1996), 5695
\item{[14]}
E.T. Newman,
J. Math. Phys. {\bf14} (1973) 102.
\item{[15]}
R.W. Lind and E.T. Newman,
J. Math. Phys. {\bf 15} (1974) 1103.
\item{[16]}
D. Ivanenko and A.Ya. Burinskii,
Izvestiya Vuzov Fiz. $N^0$ 7 (1978) 113 (Sov. Phys. J. (USA))
\item{[17]}
A. Burinsklii, R.P. Kerr, Z. Perj\'es,
Preprint gr-qc/9501012
\item{[18]}
R.P.Kerr and  W.B. Wilson,
Gen. Rel. Grav. {\bf 10} (1979) 273
\item{[19]}
A.Sen,
Phys. Rev. Lett. {\bf 69} (1992) 1006
\item{[20]}
A.Burinskii,
Phys. Rev. D {\bf 52} (1995) 5826
\end